\documentclass[11pt]{article}
\usepackage{graphicx}
\usepackage{cite}

\begin{document}

%
%
\title{\Large\bf
Exploring the structure of the proton through \\
polarization observables in {\boldmath $l \, p \to \textrm{jet} \, X$}}

\author{Zhong-Bo Kang$^{1}$, Andreas Metz$^{2}$, Jian-Wei Qiu$^{3,4}$, Jian Zhou$^{2}$
 \\[0.3cm]
{\normalsize \it $^1$RIKEN BNL Research Center, Brookhaven National Laboratory,
  Upton, NY 11973, USA} \\[0.1cm]
{\normalsize\it $^2$Department of Physics, Barton Hall,
  Temple University, Philadelphia, PA 19122, USA} \\[0.1cm]
{\normalsize\it $^3$Physics Department, Brookhaven National Laboratory, Upton, NY 11973} \\[0.1cm]
{\normalsize\it $^4$C.N.~Yang Institute for Theoretical Physics, Stony Brook University,} \\[0.1cm]
{\normalsize\it  Stony Brook, NY 11794, USA}}

\maketitle

%
%
\begin{abstract}
\noindent
We present results for a complete set of polarization observables for jet
production in lepton proton collision, where the final state lepton is not
observed.
The calculations are carried out in collinear factorization at the level
of Born diagrams.
For all the observables we also provide numerical estimates for typical
kinematics of a potential future Electron Ion Collider.
On the basis of this numerical study, the prospects for the transverse single
target spin asymmetry are particularly promising.
This observable is given by a certain quark-gluon correlation function, which
has a direct relation to the transverse momentum dependent Sivers parton
distribution.
\end{abstract}

%
%
\section{Introduction}
\noindent
In lepton nucleon scattering one normally detects the scattered lepton in
order to determine the virtuality $Q^2$ of the exchanged gauge boson.
Provided that $Q^2$ is sufficiently large, one can use, for a number of final
states, the machinery of QCD factorization (see Ref.~\cite{Collins:1989gx} for
an overview) in order to separate the short-distance physics from the
non-perturbative long-distance physics encoded in different parton correlation
functions.
In data analyses for lepton nucleon scattering, QCD factorization not only has
been used to get a handle on objects like ordinary forward parton distributions,
but also to address generalized parton distributions and transverse momentum
dependent parton distributions.

In the present work we study inclusive jet production in lepton proton
scattering with the scattered lepton going unobserved, i.e.,
$l \, p \to \textrm{jet} \, X$.
In this case, the transverse momentum of the jet can serve as the large
scale, which is needed for justifying a calculation in perturbative QCD.
The kinematics of this process is rather simple --- in particular, simpler
than the kinematics of semi-inclusive deep-inelastic scattering (DIS) ---
and in essence coincides with the one of, e.g., single jet production in
hadronic collisions.
We focus here on jet production, as opposed to hadron production, because
this process, in principle, can provide more direct information about
parton correlation functions of the proton as no uncertainties from parton
fragmentation are involved.

We consider all possible polarization observables for the process
$l \, p \to \textrm{jet} \, X$ at the level of Born diagrams.
Neglecting parity violating effects as well as transverse polarization
of the initial state lepton, one can identify three spin-dependent cross
sections: $\sigma_{LL}$, $\sigma_{UT}$, and $\sigma_{LT}$, where the first
index refers to the lepton polarization and the second one to the proton
polarization.
While $\sigma_{LL}$, like the unpolarized cross section $\sigma_{UU}$, is a
twist-2 observable, the latter two are twist-3 effects.

In order to compute $\sigma_{UT}$ and $\sigma_{LT}$ we make use of collinear
twist-3 factorization, which was pioneered already in the early
80's~\cite{Efremov:1981sh,Ellis:1982wd} and applied for the
first time to single spin asymmetries (SSAs) in~\cite{Efremov:1981sh}.
These early treatments of twist-3 SSAs were later revisited and
improved~\cite{Qiu:1991pp,Eguchi:2006mc}.
In the meantime many papers on SSAs and related observables in the collinear
twist-3 formalism exist;
see, e.g., Refs.~\cite{Ratcliffe:1998pq,Kanazawa:2000hz,Eguchi:2006qz,Kouvaris:2006zy,Koike:2006qv,Bomhof:2007su,Bacchetta:2007sz,Ratcliffe:2007ye,Qiu:2007ar,Yuan:2008it,Kang:2008qh,Zhou:2008fb,Ma:2008gm,Kang:2008ih,Koike:2009ge,Yuan:2009dw,Zhou:2009jm,Kang:2010zzb,Kanazawa:2010au,Zhou:2011ba,Kang:2011mr}.
In particular, various works are dealing with the QCD-evolution of the relevant
twist-3 correlators (see~\cite{Kang:2008ey,Zhou:2008mz,Vogelsang:2009pj,Braun:2009mi,Kang:2010xv}
and references therein).
Moreover, the relation between the collinear twist-3 factorization and a
factorization in terms of transverse momentum dependent parton
correlators~\cite{Collins:1981uk,Ji:2004wu,Collins:2004nx}
has been studied in detail for semi-inclusive deep-inelastic scattering and
the Drell-Yan process~\cite{Ji:2006ub,Koike:2007dg,Bacchetta:2008xw,Zhou:2009rp}.

Based on our numerical estimate for a typical kinematics of the currently
discussed/planned Electron Ion Collider (see, e.g.,
Ref.~\cite{DeRoeck:2009af,Anselmino:2011ay}), the transverse SSA
$A_{UT} = \sigma_{UT} / \sigma_{UU}$ appears to be rather promising.
The QCD-description of $A_{UT}$ contains a specific twist-3 quark-gluon-quark
correlator --- the so-called ETQS (Efremov-Teryaev-Qiu-Sterman) matrix
element~\cite{Efremov:1981sh,Qiu:1991pp}.
As pointed out in~\cite{Boer:2003cm,Ma:2003ut}, the ETQS matrix element is
related to the transverse momentum dependent Sivers function~\cite{Sivers:1989cc}.
(For experimental studies of the Sivers effect in semi-inclusive DIS we refer
to~\cite{Airapetian:2004tw,Alexakhin:2005iw},
while extractions of the Sivers function from data were
discussed in~\cite{Efremov:2004tp,Anselmino:2005nn,Vogelsang:2005cs,Collins:2005ie,Arnold:2008ap,Anselmino:2008sga}).
Thus, measuring $A_{UT}$ in $l \, p \to \textrm{jet} \, X$ would give a direct
(complementary) handle on the Sivers effect.
Our prediction for the longitudinal double spin asymmetry
$A_{LL} = \sigma_{LL} / \sigma_{UU}$ is at the percent level, while for $A_{LT}$,
computed in a Wandzura-Wilczek-type approximation, we obtain only a tiny effect.

Note also that a detailed study of $A_{UT}$ in $l \, p \to \textrm{jet} \, X$, based
on factorization in terms of transverse momentum dependent correlators, can be found
in Ref.~\cite{Anselmino:2009pn}.
That work represents an extension and update of a related earlier investigation of
$A_{UT}$ for $l \, p \to \pi \, X$~\cite{Anselmino:1999gd}.
Moreover, $A_{UT}$ for pion production was also considered in the collinear
twist-3 approach in two conference proceedings~\cite{Koike:2002ti}.

%
%
\section{Factorization and predictive power}
\label{sec:fac}

\noindent
In this section, we present a QCD factorization formalism for the inclusive high
transverse momentum jet production in lepton hadron collision and provide brief
arguments why this factorization formalism should be valid.
We also provide the prescription for calculating the short-distance hard parts of the
formalism, and discuss the predictive power of perturbative calculations.

%
%
\subsection{Factorization formulas}
With a large momentum transfer, the transverse momentum $P_{JT} \equiv |\vec{P}_{JT}|$
of the inclusive jet in high energy collisions, the short-distance dynamics takes place
at a time scale of $1/P_{JT}$, which is much shorter than the typical time scale of
hadronic physics, ${\cal O}({\rm fm})$.
The quantum interference taking place between these two very different time scales is
likely suppressed by the ratio of these two scales.
Perturbative QCD factorization of hadronic cross sections effectively neglects the power
suppressed quantum interference and factorizes the cross section into a product or a
convolution of two probabilities: one for finding active parton(s) inside the identified
hadron(s), and the other is the short-distance part of partonic cross section(s).
For example, the leading power contribution to single inclusive jet production at large
transverse momentum in hadron hadron collisions, $h(P) + h'(P') \to {\rm jet}(P_J) + X$,
can be factorized as~\cite{Collins:1989gx},
\begin{equation}
\frac{d\sigma^{hh'\to {\rm jet}(P_J) X}}{dP_{JT}dy}
\approx \sum_{ab} \int dx \, f_1^{a/h}(x,\mu) \int dx' f_1^{b/h'}(x',\mu)
\frac{d\hat{\sigma}^{ab\to {\rm jet}(P_J) X}}{dP_{JT}dy}(x,x',P_{JT},y,\mu) \,,
\label{eq:fac-jet}
\end{equation}
where $\sum_{ab}$ runs over all parton flavors, $f_1^{a/h}(x,\mu)$ is the (unpolarized)
parton distribution function (PDF) of flavor $a$ and momentum fraction $x$ of hadron $h$,
and $\mu$ is the factorization scale.
Since $P_{JT}$ is the only large observed momentum, Eq.~(\ref{eq:fac-jet}) is a collinear
factorization formalism~\cite{Collins:1989gx}.
The $\frac{d\hat{\sigma}^{ab\to {\rm jet}(P_J) X}}{dP_{JT}dy}(x,x',P_{JT},y,\mu)$ in
Eq.~(\ref{eq:fac-jet}) is the perturbatively calculable short-distance hard part, which is
effectively equal to the partonic jet cross section for the collision between two partons
$a$ and $b$ with all collinear sensitive contribution removed.
The predictive power of Eq.~(\ref{eq:fac-jet}) relies on our ability to calculate the
partonic hard parts systematically in perturbative QCD order-by-order in $\alpha_s$, and
the universality of PDFs.
The factorization formalism in Eq.~(\ref{eq:fac-jet}) works extremely well for describing
single inclusive jet data at the Tevatron for over ten orders of magnitude in the
production rate~\cite{Tevatron-jet}.

Our ability to calculate the short-distance partonic hard parts in Eq.~(\ref{eq:fac-jet})
relies on the fact that the factorization of short-distance dynamics is not sensitive to
the long-distance details of the colliding hadron(s).
That is, the factorization formalism in Eq.~(\ref{eq:fac-jet}), which is valid for two
colliding hadrons, should also be valid for the collision of two asymptotic partons.
By applying Eq.~(\ref{eq:fac-jet}) to the collision of two partons of various flavors,
we can derive all short-distance hard parts order-by-order in powers
of $\alpha_s$~\cite{CTEQ-hbk}.
Since the factorization formalism in Eq.~(\ref{eq:fac-jet}) is not sensitive to the
details of the colliding particles, we expect that the same factorization formalism in
Eq.~(\ref{eq:fac-jet}) is also valid for single inclusive jet production at high $P_{JT}$
in lepton-hadron collision, $l(l)+h(p)\to {\rm jet}(P_J)+X$, as
\begin{equation}
\frac{d\sigma^{lh\to {\rm jet}(P_J) X}}{dP_{JT}dy}
\approx \sum_{ab} \int dx f_1^{a/l}(x,\mu) \int dx' f_1^{b/h}(x',\mu)
\frac{d\hat{\sigma}^{ab\to {\rm Jet}(P_J) X}}{dP_{JT}dy}(x,x',P_{JT},y,\mu) \,,
\label{eq:fac-lp}
\end{equation}
where $\sum_{a}$ runs over the lepton, the photon, and all parton flavors, while $\sum_b$
runs over all parton flavors, $f_1^{a/l}(x,\mu)$ is the nonperturbative distribution to
find a lepton, photon or parton inside the colliding lepton with the momentum fraction $x$.
Like in the case of hadronic collisions, the short-distance hard part
$\frac{d\hat{\sigma}^{ab\to {\rm Jet}(P_J) X}}{dP_{JT}dy}(x,x',P_{JT},y,\mu)$ can be
perturbatively calculated order-by-order in the coupling constant by applying the
factorized formalism in Eq.~(\ref{eq:fac-lp}) to the collision between various lepton,
photon or parton states.
For the leading contribution, it might be reasonable to keep the cross section at the
lowest power in $\alpha_{em}$ while including radiative corrections from the strong
interaction in powers of $\alpha_s$.

Like all perturbative QCD factorization approaches, the predictive power of
Eq.~(\ref{eq:fac-lp}) relies on the infrared safety of the hard parts and the universality
of the long-distance distributions.
Unlike the hadronic case in Eq.~(\ref{eq:fac-jet}), the jet production in lepton hadron
collisions requires a set of new non-perturbative distributions, $f_1^{a/l}(x,\mu)$ with
$a=l,\gamma,q,\bar{q},g$.
The operator definition of the lepton PDFs should be the same as the proton PDFs' except
that the proton state is replaced by the state of the lepton \cite{Collins:1981uw}.
The operator definition of the lepton distribution inside a lepton is very similar to
that of a quark distribution,
\begin{equation}
f_1^{l'/l}(x,\mu)
= \int\frac{dy^-}{2\pi}
e^{ixP^+ \xi^-}
\langle l,\vec{s}_l|\bar{\psi}^{l'}(0)\frac{\gamma^+}{2} {\cal W}_{\gamma}(0,\xi^-)
\psi^{l'}(\xi^-)|l, \vec{s}_l\rangle \,,
\label{eq:lepton}
\end{equation}
where ${\cal W}_{\gamma}(0,\xi^-) = \exp[-ie\int_0^{\xi^-} dy^- A^+_\gamma(y^-)]$ is the
gauge link in an Abelian gauge for the lepton moving in the ``+$z$'' direction.
If we neglect the role of the strong interaction, we can calculate the lepton distribution
perturbatively in QED.
At the lowest order, $f_1^{l'/l}(x,\mu) = \delta^{l'l}\delta(1-x)$.
The operator definition of the photon distribution inside a lepton is the same as the
operator definition for the gluon distribution inside a proton with the color dependence
removed and the proton state replaced by the lepton state,
\begin{equation}
f_1^{\gamma/l}(x,\mu)
= \frac{1}{xP^+}\int\frac{d\xi^-}{2\pi}
e^{ixP^+ \xi^-}
\langle l,\vec{s}_l|
F^{+\alpha}_{\rm em}(0)F^{+\beta}_{\rm em}(\xi^-)|l, \vec{s}_l\rangle (-g_{\alpha\beta}) \,,
\label{eq:photon}
\end{equation}
where $F^{\mu\nu}_{\rm em}$ is the electromagnetic field strength tensor, which can be
expressed in terms of gauge invariant electric and magnetic fields.
The photon distribution of the lepton, $f_1^{\gamma/l}(x,\mu)$, could be calculated
perturbatively in QED if we neglect the strong interaction.
But, in general, it is a non-perturbative distribution.
\begin{figure}[t]
\begin{center}
\includegraphics[width=7.5cm]{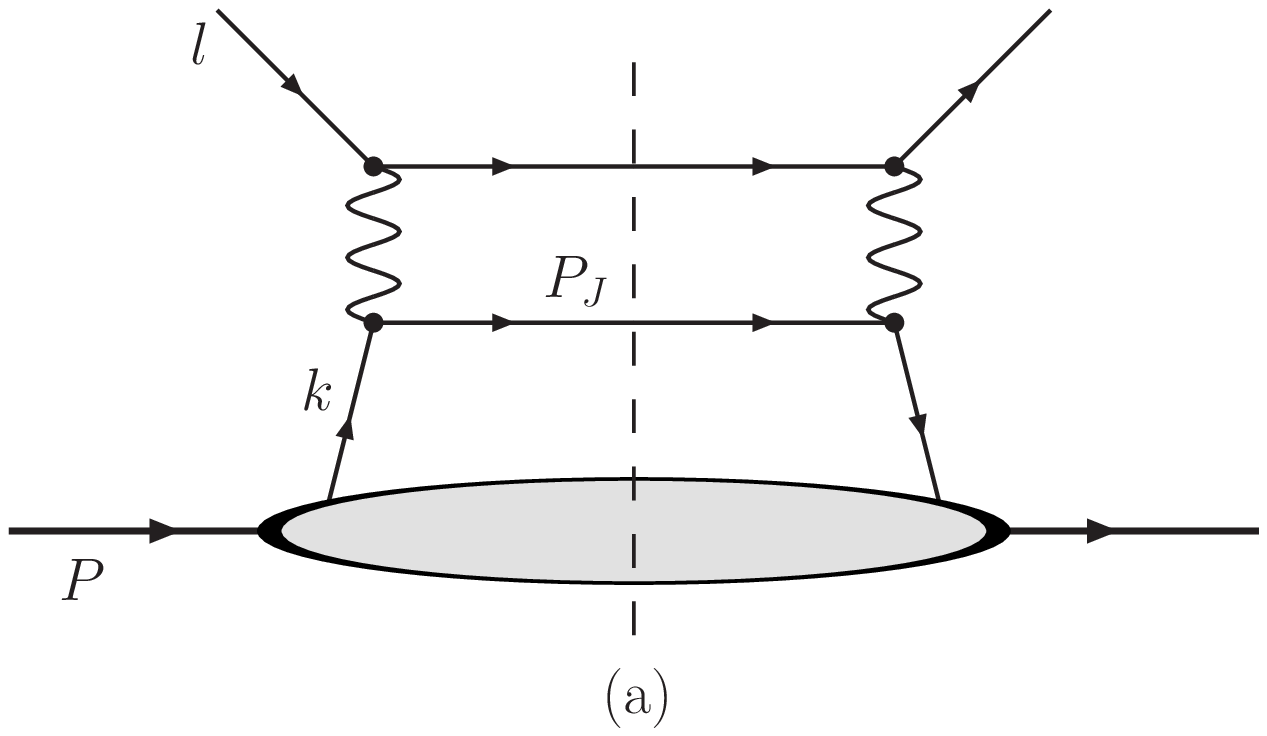}
\hskip 1.0cm
\includegraphics[width=7.5cm]{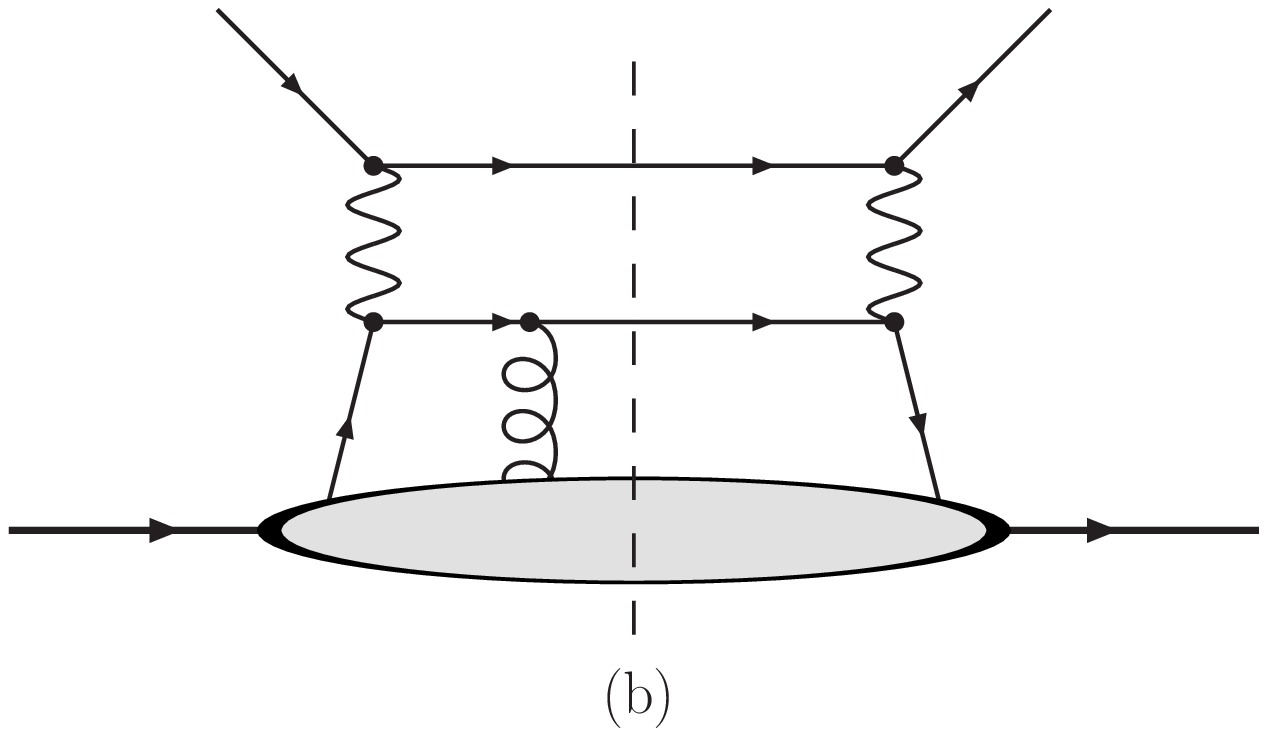}
\caption{Diagram (a): parton model representation for $l \, p \to \textrm{jet} \, X$.
The jet is produced by the struck quark, and the final state lepton goes unobserved.
Diagram (b): contribution from quark-gluon-quark correlation. This diagram, together
with its Hermitian conjugate which is not displayed, needs to be taken into account
when computing twist-3 observables.}
\label{f:diagram}
\end{center}
\end{figure}

Based on the same argument that the factorization of the short-distance dynamics is not
sensitive to the details of the colliding particles, and the fact that the single inclusive
jet cross section at high transverse momentum has only one observed large momentum transfer
$P_{JT}$, we expect that the twist-3 collinear factorization formalism for calculating
the transverse-spin dependent cross section in hadronic
collisions~\cite{Efremov:1981sh,Qiu:1991pp,Eguchi:2006mc} can be applied to the single
inclusive jet cross section in lepton hadron collisions.
Therefore,
\begin{equation}
\frac{d\Delta\sigma^{lh\to {\rm jet}(P_J) X}({S}_T)}{dP_{JT}dy}
\approx \sum_{ab} \int dx f_1^{a/l}(x,\mu) \int dx' T_F^{b/h}(x',x',\mu,{S}_T)
H^{ab\to {\rm Jet}(P_J) X}(x,x',P_{JT},y,\mu) \,,
\label{eq:fac3-lp}
\end{equation}
where $\Delta\sigma({S}_T)\equiv [\sigma({S}_T)-\sigma(-{S}_T)]/2$ is the transverse-spin
dependent cross section, with the spin vector $S_T \equiv |\vec{S}_T|$ of the transversely
polarized colliding hadron.
The $T_F^{b/h}(x,x,\mu,\vec{S}_T)$ with $b=q,\bar{q},g$ in Eq.~(\ref{eq:fac3-lp}) is the
universal twist-3 parton correlation function relevant for the SSAs \cite{Qiu:1991pp,Kang:2008ey},
and the $H^{ab\to {\rm Jet}(P_J) X}(x,x',P_{JT},y,\mu)$ with $a=l,\gamma,q,\bar{q},g$ and
$b=q,\bar{q},g$ are the process-dependent short-distance hard parts whose leading order
contributions are derived below.

%
%
\subsection{Partonic hard parts}
In this subsection, we provide the prescription for calculating the partonic hard parts of
the factorization formulas in Eqs.~(\ref{eq:fac-lp}) and~(\ref{eq:fac3-lp}).

To calculate the short-distance hard parts,
$\frac{d\hat{\sigma}^{ab\to {\rm Jet}(P_J) X}}{dP_{JT}dy}(x,x',P_{JT},y,\mu)$ in
Eq.~(\ref{eq:fac-lp}), we apply the factorization formalism to the collision between all
possible combinations of two asymptotic incoming lepton, photon, or parton states.
For the leading order (LO) contribution, the jet cross section is given by the lowest order
lepton-quark scattering, as sketched in Fig.~\ref{f:diagram}(a), and the jet is effectively
given by the final-state quark: ${\rm jet}(P_J)\to q(P_J)$ at the lowest order.
The corresponding hard part can be uniquely derived by applying Eq.~(\ref{eq:fac-lp}) to
the collision of the lepton on a quark state: $l\to l$ and $h\to q$,
\begin{equation}
\sigma_{(2,0)}^{lq\to q(P_J)\to {\rm jet}(P_J)} =
f_{1(0)}^{l/l} \otimes f_{1(0)}^{q/q} \otimes
\hat{\sigma}_{(2,0)}^{lq\to q(P_J)\to{\rm jet}(P_J)} =
\hat{\sigma}_{(2,0)}^{lq\to q(P_J)\to{\rm jet}(P_J)} \,,
\label{eq:lq-lo}
\end{equation}
where the superscript ``(2,0)'' indicates two powers of $\alpha_{em}$ and zeroth order in
$\alpha_s$, and $\otimes$ represents the convolution over lepton or parton momentum fraction
as shown in Eq.~(\ref{eq:fac-lp}).
At the lowest order, the short-distance hard part for the jet cross section is effectively
the same as the lepton-quark scattering cross section for producing the quark $q(P_J)$ and
is perturbatively finite.
\begin{figure}[t]
\begin{center}
\begin{minipage}[t]{5cm}
\begin{center}
\includegraphics[width=2.0cm]{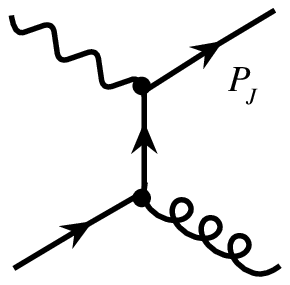}
\hskip 0.5cm
\includegraphics[width=2.0cm]{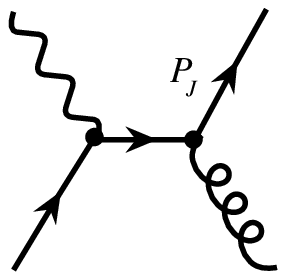}
\\
(a)
\end{center}
\end{minipage}
\hskip 2.0cm
\begin{minipage}[t]{5.8cm}
\begin{center}
\includegraphics[width=2.4cm]{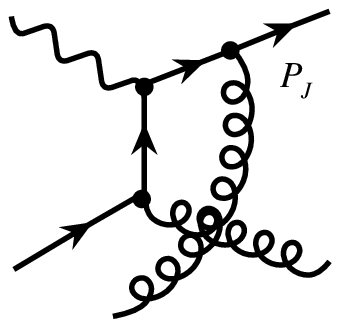}
\hskip 0.5cm
\includegraphics[width=2.4cm]{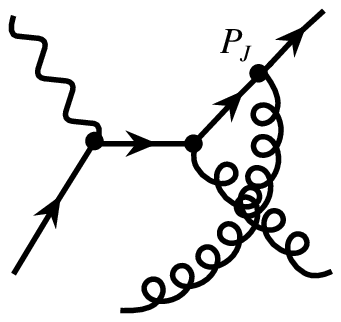}
\\
(b)
\end{center}
\end{minipage}
\caption{(a) Leading order tree diagrams for photon scattering on an asymptotic quark state.
(b) Leading order tree diagrams for photon scattering on an asymptotic quark-gluon composite
state. Note that only the gluon interaction with the observed final-state parton is nonzero,
while the interaction with the unobserved parton cancels.}
\label{f:photon-quark}
\end{center}
\end{figure}

If we apply the factorization formalism in Eq.~(\ref{eq:fac-lp}) to photon-quark collision
by letting $l \to \gamma$ and $h\to q$, as sketched in Fig.~\ref{f:photon-quark}(a), we can
have two additional tree-level contributions to the jet cross section,
\begin{eqnarray}
\sigma_{(1,1)}^{\gamma q\to q(P_J)\to {\rm jet}(P_J)}
=
f_{1(0)}^{\gamma/\gamma} \otimes f_{1(0)}^{q/q} \otimes
\hat{\sigma}_{(1,1)}^{\gamma q\to q(P_J)\to{\rm jet}(P_J)}
=
\hat{\sigma}_{(1,1)}^{\gamma q\to q(P_J)\to{\rm jet}(P_J)} \,,
\nonumber \\
\sigma_{(1,1)}^{\gamma q\to g(P_J)\to {\rm jet}(P_J)}
=
f_{1(0)}^{\gamma/\gamma} \otimes f_{1(0)}^{q/q} \otimes
\hat{\sigma}_{(1,1)}^{\gamma q\to g(P_J)\to{\rm jet}(P_J)}
=
\hat{\sigma}_{(1,1)}^{\gamma q\to g(P_J)\to{\rm jet}(P_J)} \,.
\label{eq:gq-lo}
\end{eqnarray}
The difference of these two contributions is whether the jet is generated by an energetic
quark or a gluon.
Similarly, if we apply Eq.~(\ref{eq:fac-lp}) to photon-gluon collision, we obtain two more
tree-level contributions to the jet cross section:
$\hat{\sigma}_{(1,1)}^{\gamma g\to q(P_J)\to {\rm jet}(P_J)}$ and
$\hat{\sigma}_{(1,1)}^{\gamma g\to \bar{q}(P_J)\to {\rm jet}(P_J)}$.
Since the photon distribution of the lepton $f_1^{\gamma/l}(x,\mu)$ carries at least one power
of $\alpha_{em}$ higher than the leading term of $f_1^{l/l}(x,\mu)$, these photon-parton
contributions could be formally considered as an order $\alpha_s$ correction to the LO term
in Eq.~(\ref{eq:lq-lo}).
However, since the photon distribution of the lepton $f_1^{\gamma/l}(x,\mu)$ has a large
QED logarithm perturbatively, for certain kinematics these terms could be more important than
typical higher order corrections.

If we apply the factorization formalism in Eq.~(\ref{eq:fac-lp}) to parton-parton collision
by letting $l \to a$ and $h\to b$ with $a,b=q,\bar{q},g$, the partonic hard parts are to be
the same as those in the hadronic collisions \cite{Qiu:1991pp,Kouvaris:2006zy}.
Since the parton distribution functions of the lepton $f_1^{a/l}(x,\mu)$ have at least two
powers of $\alpha_{em}$ perturbatively, even the leading Born contribution here should be
considered as higher order correction in powers of $\alpha_s$.

\begin{figure}[t]
\begin{center}
\begin{minipage}[t]{6.2cm}
\begin{center}
\includegraphics[width=2.6cm]{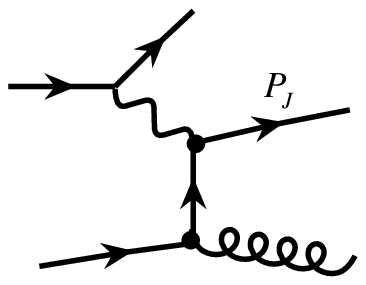}
\hskip 0.5cm
\includegraphics[width=2.6cm]{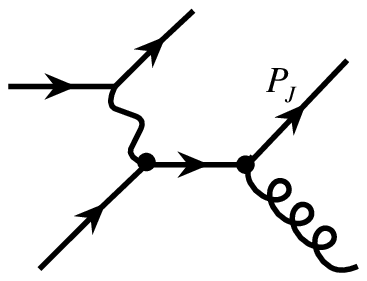}
\\
(a)
\end{center}
\end{minipage}
\hskip 2.0cm
\begin{minipage}[t]{6.2cm}
\begin{center}
\includegraphics[width=2.6cm]{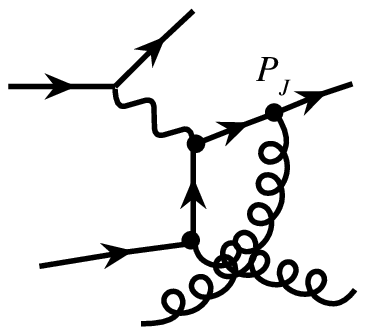}
\hskip 0.5cm
\includegraphics[width=2.6cm]{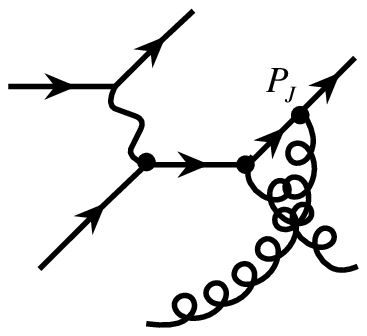}
\\
(b)
\end{center}
\end{minipage}
\caption{(a) Next-to-leading order tree diagrams for lepton scattering on an asymptotic quark state.
(b) Next-to-leading order tree diagrams for lepton scattering on an asymptotic quark-gluon
composite state.}
\label{f:lepton-quark2}
\end{center}
\end{figure}

Higher order corrections to the short-distance hard parts of the factorization formalism can
be derived in the same way by applying Eq.~(\ref{eq:fac-lp}) to the scattering of various
partonic states at higher orders in $\alpha_s$.
For example, we can calculate the next-to-leading order (NLO) contribution
$\hat{\sigma}_{(2,1)}^{lq\to {\rm jet}(P_J)}$ by applying Eq.~(\ref{eq:fac-lp}) to the lepton-quark
collision at order $\alpha_s$,
\begin{eqnarray}
\sigma_{(2,1)}^{lq\to {\rm jet}(P_J)}
&=&
f_{1(0)}^{l/l} \otimes f_{1(0)}^{q/q} \otimes \hat{\sigma}_{(2,1)}^{lq\to {\rm jet}(P_J)} +
f_{1(0)}^{l/l} \otimes f_{1(1)}^{q/q} \otimes \hat{\sigma}_{(2,0)}^{lq\to {\rm jet}(P_J)}
\nonumber\\
&\ &
+f_{1(1)}^{\gamma/l} \otimes f_{1(0)}^{q/q} \otimes
\Big[ \hat{\sigma}_{(1,1)}^{\gamma q\to q(P_J)}
    + \hat{\sigma}_{(1,1)}^{\gamma q\to g(P_J)} \Big] \,,
\label{eq:nlo}
\end{eqnarray}
which can be written as
\begin{eqnarray}
\hat{\sigma}_{(2,1)}^{lq\to {\rm jet}(P_J)}
&=&
\sigma_{(2,1)}^{lq\to {\rm jet}(P_J)}
-f_{1(1)}^{q/q} \otimes \hat{\sigma}_{(2,0)}^{lq\to {\rm jet}(P_J)}
-f_{1(1)}^{\gamma/l} \otimes \Big[ \hat{\sigma}_{(1,1)}^{\gamma q\to q(P_J)}
     + \hat{\sigma}_{(1,1)}^{\gamma q\to g(P_J)} \Big] \,,
\label{eq:lq-nlo}
\end{eqnarray}
where the first term on the right-hand-side (RHS) is the partonic cross section given by the
real Feynman diagrams sketched in Fig.~\ref{f:lepton-quark2}(a) plus the virtual diagrams
from one-loop corrections to the tree-diagram in Fig.~\ref{f:diagram}(a).
The real diagrams in Fig.~\ref{f:lepton-quark2}(a) have three potential collinear logarithmic divergences.  One comes from the final-state gluon radiation when the gluon is parallel to the parent quark.  Such final-state collinear divergence is taken care of by the jet definition and its finite cone size.  The other two come from the initial-state:
when the gluon line is almost parallel to the incoming quark or when the photon is about parallel to the incoming lepton.  As a result of QCD factorization, these two divergences are systematically removed by the second and the third terms in Eq.~(\ref{eq:lq-nlo}), respectively.
The hard $2\to 2$ scattering cross section of the second term,
$\hat{\sigma}_{(2,0)}^{lq\to {\rm jet}(P_J)}$ is derived in Eq.~(\ref{eq:lq-lo}), while
$\hat{\sigma}_{(1,1)}^{\gamma q\to q(P_J)}$ and
$\hat{\sigma}_{(1,1)}^{\gamma q\to g(P_J)}$ of the third term are given in
Eq.~(\ref{eq:gq-lo}).

Similarly, we can derive another short-distance contribution at the same order,
$\hat{\sigma}_{(2,1)}^{l g\to {\rm jet}(P_J)}$, by applying Eq.~(\ref{eq:fac-lp}) to the
collision between a lepton and a gluon,
\begin{eqnarray}
\hat{\sigma}_{(2,1)}^{lg\to {\rm jet}(P_J)}
&=&
\sigma_{(2,1)}^{lg\to {\rm jet}(P_J)}
-f_{1(1)}^{q/g} \otimes \hat{\sigma}_{(2,0)}^{lq\to {\rm jet}(P_J)}
-f_{1(1)}^{\gamma/l} \otimes
\Big[ \hat{\sigma}_{(1,1)}^{\gamma g\to q(P_J)}
+ \hat{\sigma}_{(1,1)}^{\gamma g\to \bar{q}(P_J)} \Big] \,,
\label{eq:lg-nlo}
\end{eqnarray}
where the second and the third terms on the RHS again remove the collinear divergence of
the partonic scattering cross section.
In general, the perturbatively calculated hard parts are effectively equal to the partonic
cross sections with all collinear divergences removed.

In the collinear factorization approach, the spin-dependent cross section to the SSAs
comes from the interference of the real part of the scattering amplitude with one active
parton and the imaginary part of the scattering amplitude with two active
partons~\cite{Efremov:1981sh,Qiu:1991pp,Eguchi:2006mc,Kouvaris:2006zy}.
For the LO contribution, the partonic hard part $H^{lq\to q(P_J) X}(x,x',P_{JT},y)$ in
Eq.~(\ref{eq:fac3-lp}) is given by the diagram in Fig.~\ref{f:diagram}(b), and will be
calculated in the following section.

For the higher order corrections, the partonic hard part
$H^{ab\to {\rm Jet}(P_J) X}(x,x',P_{JT},y,\mu)$ in Eq.~(\ref{eq:fac3-lp}) can be calculated
in the same way by applying the formalism to the various partonic states.
For example, the LO photon-parton scattering contribution to the SSA comes from the
interference of Feynman diagrams in Fig.~\ref{f:photon-quark}(a) and \ref{f:photon-quark}(b).
Similarly, the NLO lepton-quark scattering contribution to the SSA comes from the interference
of Feynman diagrams in Fig.~\ref{f:lepton-quark2}(a) and \ref{f:lepton-quark2}(b).
In the rest of this paper, we will present our derivation and results of the LO contribution
to various spin asymmetries.
We will leave the explicit treatment of higher order corrections to future work.

%
%
\section{Kinematics and analytical results}
\noindent
In this section we present some details of the kinematics for the process
$l(l) + p(P) \to \textrm{jet}(P_J) + X$, as well as the tree level formulas for
the various observables.
We use the momenta of the particles to fix a coordinate system according to
$\hat{e}_z = \vec{P}/|\vec{P}| = - \vec{l}/|\vec{l}|$,
$\hat{e}_x = \vec{P}_{JT}/|\vec{P}_{JT}|$, and $\hat{e}_y = \hat{e}_z \times \hat{e}_x$.
Mandelstam variables are defined by
\begin{equation} \label{e:mandel_1}
s = (l + P)^2 \,, \qquad
t = (P - P_J)^2 \,, \qquad
u = (l - P_J)^2 \,,
\end{equation}
while on the partonic level one has
\begin{equation} \label{e:mandel_2}
\hat{s} = (l + k)^2 = x s\,, \qquad
\hat{t} = (k - P_J)^2 = x t\,, \qquad
\hat{u} = (l - P_J)^2 = u \,,
\end{equation}
with $k$ denoting the momentum of the active quark in the proton;
see also Fig.~\ref{f:diagram}(a).
The momentum fraction $x$ specifies the plus-momentum of the quark through
$k^+ = x P^+$.\footnote{For a generic 4-vector $v$, we define light-cone coordinates
according to $v^{\pm} = (v^0 \pm v^3) / \sqrt{2}$ and $\vec{v}_T = (v^1,v^2)$.}
Using $\hat{s} + \hat{t} + \hat{u} = 0$ one finds that $x = -u/(s+t)$.
In other words, the longitudinal momentum of the struck quark is fixed by
the external kinematics of the process, like it is in fully inclusive DIS.
Of course, this no longer applies once higher order corrections are taken into
account.
For the numerical estimates we will use $P_{JT}$, and the Feynman variable $x_F$
(defined in the lepton-proton {\it cm}-frame) for which one has
\begin{equation}
x_F = \frac{2 P_{Jz}}{\sqrt{s}} = \frac{t - u}{s} \,.
\end{equation}

Next, we turn to the polarization observables for $l \, p \to \textrm{jet} \, X$,
which we compute in the collinear factorization framework.
We restrict ourselves to one-photon exchange between the leptonic and the hadronic
part of the process.
Allowing for longitudinal polarization of the initial state lepton, as well as
longitudinal and transverse polarization of the proton target, one finds
the following expression for the cross
section:\footnote{Polarization degrees are suppressed in the cross section
formula~(\ref{e:master}).}
\begin{eqnarray} \label{e:master}
P_J^0 \frac{d^3 \sigma}{d^3 P_J} & = &
\frac{\alpha_{em}^2}{s} \sum_a \frac{e_a^2}{(s+t) \, x} \, \bigg\{
f_1^a(x) \, H_{UU} +
\lambda_l \lambda_p \, g_1^a(x) \, H_{LL}
\nonumber \\
&& \hspace{2.0cm}
 + \, 2 \pi M \, \varepsilon_T^{ij} S_T^i P_{JT}^j \,
 \bigg[ \, T_{F}^a(x,x) - x \, \frac{d}{dx}T_{F}^a(x,x) \bigg]
 \frac{\hat{s}}{\hat{t} \hat{u}} \, H_{UU}
\nonumber \\
&& \hspace{2.0cm}
 + \, \lambda_l \, 2M \, \vec{S}_T \cdot \vec{P}_{JT} \,
 \bigg[ \bigg( \tilde{g}^a(x) - x \frac{d}{dx} \tilde{g}^a(x) \bigg)
 \frac{\hat{s}}{\hat{t} \hat{u}} \, H_{LL} +
 x \, g_T^a(x) \, \frac{2}{\hat{t}} \bigg]
\bigg\} \,.
\end{eqnarray}
In Eq.~(\ref{e:master}), which is the main analytical result of our work,
$\lambda_l$ and $\lambda_p$ represent the helicity of the lepton and the proton,
respectively.
One can project out the four independent components of the cross section
in~(\ref{e:master}) according to
\begin{eqnarray} \label{e:sigma_uu}
\sigma_{UU} & = & \frac{1}{4}
 \Big(\sigma(+,+) + \sigma(-,+) + \sigma(+,-) + \sigma(-,-) \Big) \,,
\\ \label{e:sigma_ll}
\sigma_{LL} & = & \frac{1}{4}
 \Big( \big[ \sigma(+,+) - \sigma(-,+) \big] - \big[ \sigma(+,-) - \sigma(-,-) \big] \Big) \,,
\\ \label{e:sigma_ut}
\sigma_{UT} & = & \frac{1}{4}
 \Big( \big[ \sigma(+,\uparrow_{y}) + \sigma(-,\uparrow_{y}) \big]
     - \big[ \sigma(+,\downarrow_{y}) + \sigma(-,\downarrow_{y}) \big] \Big) \,,
\\ \label{e:sigma_lt}
\sigma_{LT} & = & \frac{1}{4}
 \Big( \big[ \sigma(+,\uparrow_{x}) - \sigma(-,\uparrow_{x}) \big]
     - \big[ \sigma(+,\downarrow_{x}) - \sigma(-,\downarrow_{x}) \big] \Big) \,.
\end{eqnarray}
In these formulas, '$+$' and '$-$' indicate particle helicities, whereas
'$\uparrow_{x/y}$' ('$\downarrow_{x/y}$') denotes transverse polarization of the proton
along $\hat{e}_{x/y} \, (-\hat{e}_{x/y})$.

As already mentioned, both $\sigma_{UU}$ and $\sigma_{LL}$ are twist-2 observables.
We computed them on the basis of diagram (a) in Fig.~\ref{f:diagram} by applying the
collinear approximation to the momentum $k$ of the active quark.
In the case of $\sigma_{UU}$ the result contains the unpolarized quark distribution
$f_1^a$, while for $\sigma_{LL}$ the quark helicity distribution $g_1^a$ shows up.
The hard scattering coefficients for these two terms in~(\ref{e:master}), expressed through
the partonic Mandelstam variables in~(\ref{e:mandel_2}), read
\begin{equation}
H_{UU} = \frac{2 (\hat{s}^2 + \hat{u}^2)}{\hat{t}^2} \,, \qquad
H_{LL} = \frac{2 (\hat{s}^2 - \hat{u}^2)}{\hat{t}^2} \,.
\end{equation}

The cross sections $\sigma_{UT}$ and $\sigma_{LT}$ (3rd and 4th term on the
r.h.s.~of~(\ref{e:master}), respectively) represent twist-3 observables.
(Note that $M$ is the proton mass, and $\varepsilon_T^{ij} \equiv \varepsilon^{-+ij}$
with $\varepsilon^{0123} = 1$.)
The transverse SSA $A_{UT} = \sigma_{UT} / \sigma_{UU}$ is analogous to the SSA $A_N$
which has been extensively studied in one-particle inclusive production for hadron-hadron
collisions; see also Ref.~\cite{Anselmino:2009pn},
and~\cite{Adams:2003fx,Adler:2005in,Arsene:2008mi} for experimental results from RHIC.
A similar observable was proposed in~\cite{She:2008tu,Sun:2009ew} for semi-inclusive DIS,
but in this case the final state lepton still needs to be observed.

Calculational details for such twist-3 observables in collinear factorization can be
found in various papers; see, e.g.,
Refs.~\cite{Qiu:1991pp,Eguchi:2006mc,Kouvaris:2006zy,Zhou:2009jm}.
We merely mention that one has to expand the hard scattering contributions around
vanishing transverse parton momenta.
While for twist-2 effects only the leading term of that expansion matters, in the case
of twist-3 the second term is also relevant.
In addition, the contribution from quark-gluon-quark correlations, as displayed in
diagram (b) in Fig.~\ref{f:diagram}, needs to be taken into consideration.
The sum of all the terms can be written in a color gauge invariant form, which provides
a consistency check of the calculation.

The quark-gluon-quark correlator showing up in $\sigma_{UT}$ is the aforementioned
ETQS matrix element $T_F^a(x,x)$~\cite{Efremov:1981sh,Qiu:1991pp}.
The peculiar feature of this object is the vanishing gluon momentum --- that's why
it is also called ``soft gluon pole matrix element''.
If the gluon momentum becomes soft one can hit the pole of a quark propagator in
the partonic scattering
process, providing an imaginary part (nontrivial phase) which,
quite generally, can lead to single spin
effects~\cite{Efremov:1981sh,Qiu:1991pp}.
Note also that in our lowest order calculation of $\sigma_{UT}$ no so-called soft
fermion pole contribution (see~\cite{Koike:2009ge} and references therein) emerges.
For $\sigma_{LT}$ another quark-gluon-quark matrix element --- denoted as
$\tilde{g}^a$; see, in particular,
Refs.~\cite{Eguchi:2006qz,Zhou:2008mz,Zhou:2009jm} --- appears, together with the
familiar twist-3 quark-quark correlator $g_T^a$.

We use the common definitions for $g_1$ and $g_T$.
The quark-gluon-quark correlators $T_F$ and $\tilde{g}$ are specified according
to\footnote{Note that in the literature different conventions for $T_F$ exist.}
\begin{eqnarray} \label{e:defqgq_1}
T_F(x,x) & = &
\frac{1}{2M} \int \frac{d\xi^- d\zeta^-}{(2\pi)^2} \, e^{i x P^+ \xi^-} \,
\langle P,S_T | \bar{\psi}(0) \, \gamma^+ \,
              ig F^{+i}(\zeta^-) \, \psi(\xi^-) | P,S_T \rangle
              \left(i\, \varepsilon_T^{ij} S_T^j\right) \,,
\\ \label{e:defqgq_2}
\tilde{g}(x) & = &
\frac{1}{2M} \int \frac{d\xi^-}{2\pi} \, e^{i x P^+ \xi^-} \,
\left(S_T^i \right)
\nonumber \\
&& \hspace{1.0cm} \mbox{} \times
\langle P,S_T | \bar{\psi}(0) \, \gamma_5 \gamma^+ \,
              \bigg( i D_T^i - ig \int_0^\infty d\zeta^- F^{+i}(\zeta^-) \bigg) \,
              \psi(\xi^-) | P,S_T \rangle \,,
\end{eqnarray}
with $F^{\mu\nu}$ representing the gluon field strength tensor, and
$D^{\mu} = \partial^{\mu} - i g A^{\mu}$ the covariant derivative.
Equations~(\ref{e:defqgq_1}) and~(\ref{e:defqgq_2}) hold in the light-cone gauge
$A^+ = 0$, while in a general gauge Wilson lines need to be inserted between the
field operators.

It is important that $T_F$ and $\tilde{g}$ are related to moments of transverse
momentum dependent parton distributions.
To be explicit, one has~\cite{Boer:2003cm,Ma:2003ut,Zhou:2008mz,Zhou:2009jm}
\begin{eqnarray} \label{e:qgq_tmd_1}
\pi \, T_F(x,x) & = &
- \int d^2k_T \, \frac{\vec{k}_T^2}{2M^2} \,
              f_{1T}^{\perp}(x,\vec{k}_T^2)\Big|_{DIS} \,,
\\ \label{e:qgq_tmd_2}
\tilde{g}(x) & = &
\int d^2k_T \, \frac{\vec{k}_T^2}{2M^2} \, g_{1T}(x,\vec{k}_T^2) \,,
\end{eqnarray}
where we use the conventions of Refs.~\cite{Mulders:1995dh,Boer:1997nt,Bacchetta:2006tn}
for the transverse momentum dependent correlators $f_{1T}^\perp$ and $g_{1T}$.
In Eq.~(14) we take into account that the Sivers function
$f_{1T}^\perp$~\cite{Sivers:1989cc} depends on the process in which it is
probed~\cite{Collins:2002kn,Brodsky:2002rv}.
In order to obtain numerical estimates for $\sigma_{UT}$ and $\sigma_{LT}$ we will
exploit the relations in~(\ref{e:qgq_tmd_1}), (\ref{e:qgq_tmd_2}).

%
%
\section{Numerical estimates}
\noindent
Now we move on to discuss numerical estimates for the polarization observables.
To this end we consider the three spin asymmetries $A_{LL}$, $A_{UT}$, and $A_{LT}$,
whose definitions are repeated here for convenience,
\begin{equation} \label{e:asymm}
A_{LL} = \frac{\sigma_{LL}}{\sigma_{UU}} \,, \qquad
A_{UT} = \frac{\sigma_{UT}}{\sigma_{UU}} \,, \qquad
A_{LT} = \frac{\sigma_{LT}}{\sigma_{UU}} \,.
\end{equation}
To compute $\sigma_{UU}$ we use the unpolarized parton distributions
from the CTEQ5-parameterization~\cite{Pumplin:2002vw}. The helicity
distributions entering $\sigma_{LL}$ are taken from the
GRSV-parameterization~\cite{Gluck:2000dy}. For the ETQS matrix
element $T_F$ we explore two choices: (1) we use the
relation~(\ref{e:qgq_tmd_1}) between $T_F$ and the Sivers function,
and take $f_{1T}^\perp$ from the recent fit provided in
Ref.~\cite{Anselmino:2008sga}; (2) we use $T_F$ from the
parameterization obtained in~\cite{Kouvaris:2006zy}
--- taking into consideration the recently discovered sign
change~\cite{Kang:2011hk} --- by fitting transverse SSAs measured in hadronic
collisions.

In the case of $\sigma_{LT}$ one needs input for $g_T$ and $\tilde{g}$.
For $g_T$ we resort to the frequently used Wandzura-Wilczek
approximation~\cite{Wandzura:1977qf} (see~\cite{Accardi:2009au} for a recent study
of the quality of this approximation)
\begin{equation} \label{e:appr_1}
g_T(x) \approx \int_x^1 \frac{dy}{y} \, g_1(y) \,,
\end{equation}
whereas for $\tilde{g}$ we use~(\ref{e:qgq_tmd_2}) and a Wandzura-Wilczek-type
approximation for the particular $k_T$-moment of $g_{1T}$
in~(\ref{e:qgq_tmd_2})~\cite{Metz:2008ib}, leading to
\begin{equation} \label{e:appr_2}
\tilde{g}(x) \approx x \int_x^1 \frac{dy}{y} \, g_1(y) \,.
\end{equation}
We mention that~(\ref{e:appr_2}) and a corresponding relation between chiral-odd
parton distributions were used in~\cite{Kotzinian:2006dw,Avakian:2007mv} in order
to estimate certain spin asymmetries in semi-inclusive DIS.
The comparison to data discussed in~\cite{Avakian:2007mv} looks promising, though
more experimental information is needed for a thorough test of approximate
relations like the one in~(\ref{e:appr_2}).
Measuring the double spin asymmetry $A_{LT}$, which we consider in the present
paper, may provide such a test.

For our numerical estimates we use leading order parton distributions, and
take into account the three light quark flavors.
The transverse momentum of the jet $P_{JT}$ serves as the scale for the
parton distributions.
For the following reasons the scale-dependence of all the asymmetries is rather
weak:
because the leading order evolution kernels for $f_1$ and $g_1$ are the same,
$A_{LL}$ is almost scale-independent.
This also applies to $A_{LT}$ when using the approximations~(\ref{e:appr_1})
and~(\ref{e:appr_2}).
Since both the parameterization of the Sivers function in~\cite{Anselmino:2008sga}
as well as the one for the ETQS matrix element $T_F$ in~\cite{Kouvaris:2006zy,Kang:2011hk}
are related to the unpolarized distribution $f_1$, also the scale dependence of $A_{UT}$
is quite mild.

Our results are for typical kinematics accessible at a potential future Electron
Ion Collider~\cite{DeRoeck:2009af,Anselmino:2011ay}: we consider the energies
$\sqrt{s} = 50 \, \textrm{GeV}$ and $\sqrt{s} = 100 \, \textrm{GeV}$.
The asymmetries are either presented as function of $x_F$ for fixed $P_{JT}$ or
vice versa.
\begin{figure}[t]
\begin{center}
\includegraphics[width=7.5cm]{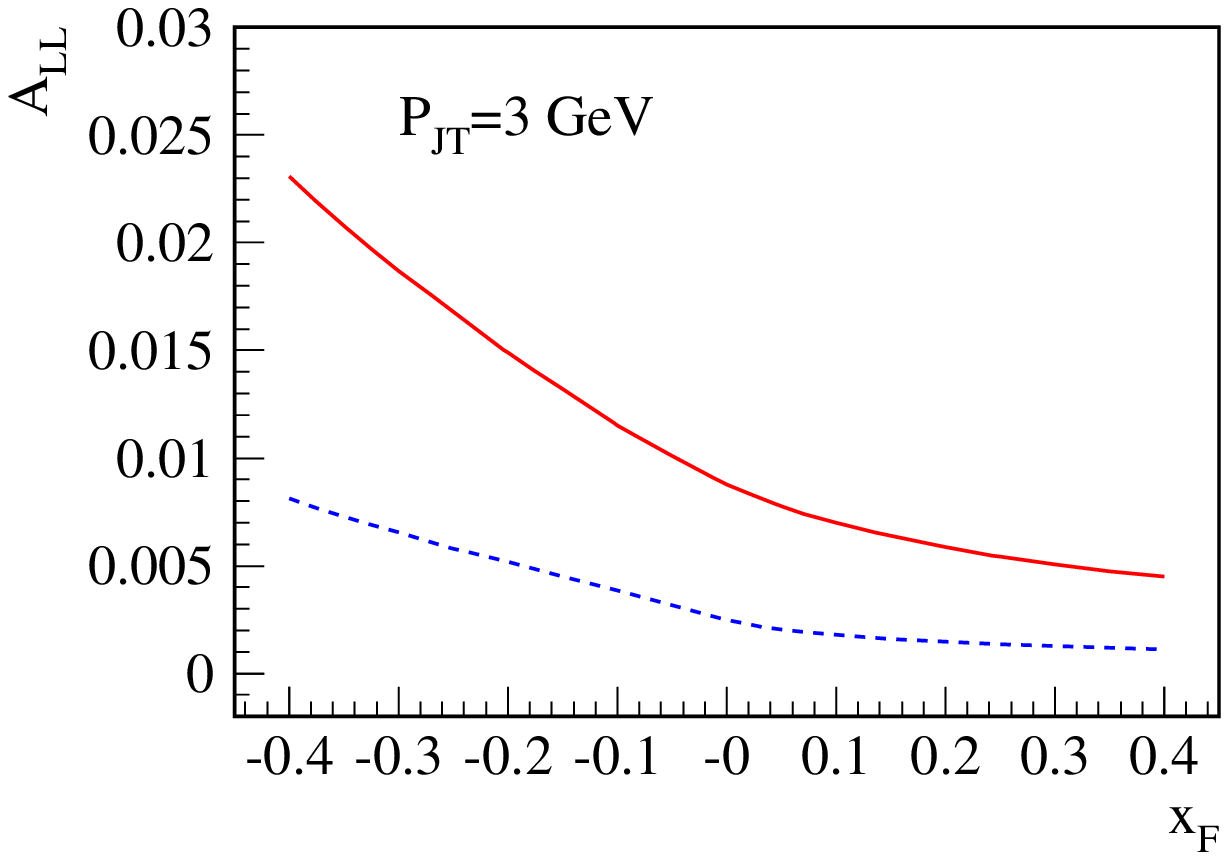}
\hskip 1.0cm
\includegraphics[width=7.5cm]{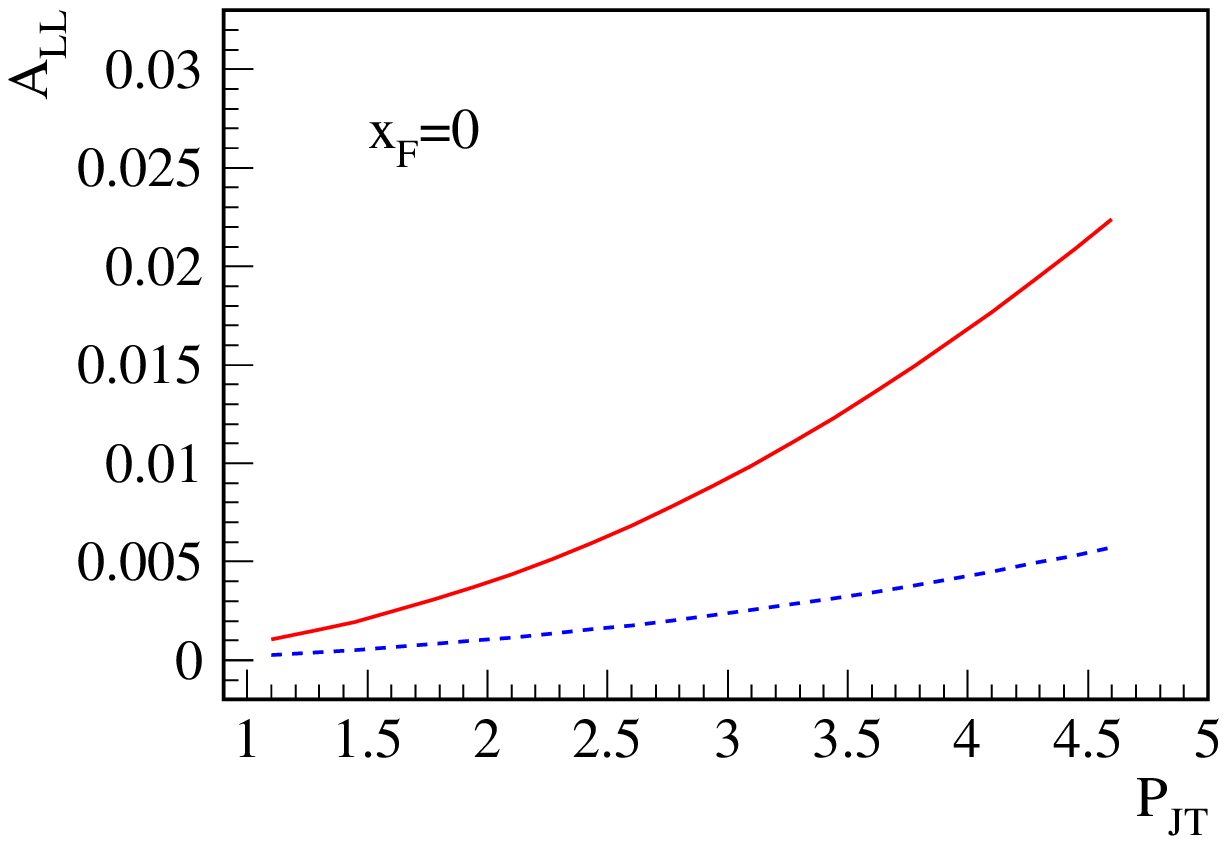}
\caption{$A_{LL}$ as a function of $x_F$ (left) and $P_{JT}$ (right).
Solid line: $\sqrt{s} = 50 \, \textrm{GeV}$;
dashed line: $\sqrt{s} = 100 \, \textrm{GeV}$.}
\label{f:all}
\end{center}
\end{figure}
\begin{figure}[t]
\begin{center}
\includegraphics[width=7.5cm]{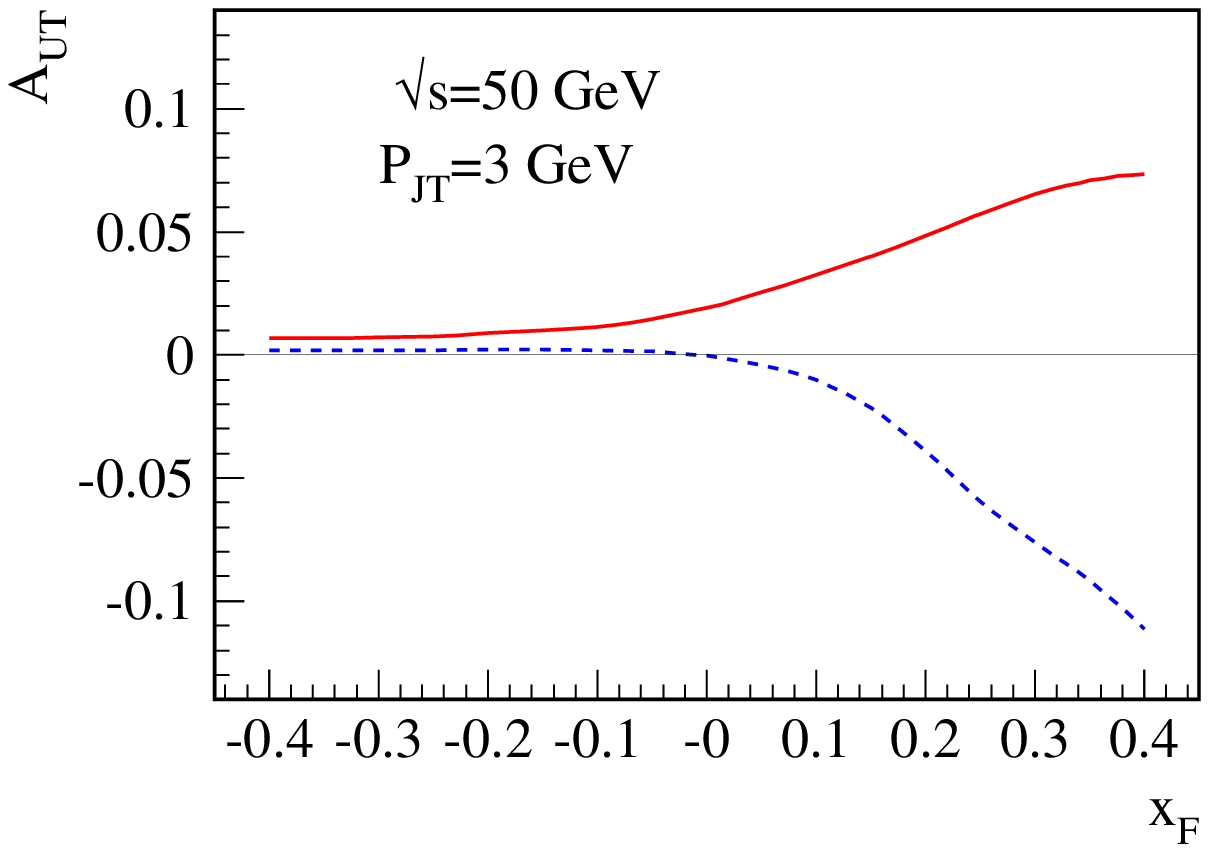}
\hskip 1.0cm
\includegraphics[width=7.5cm]{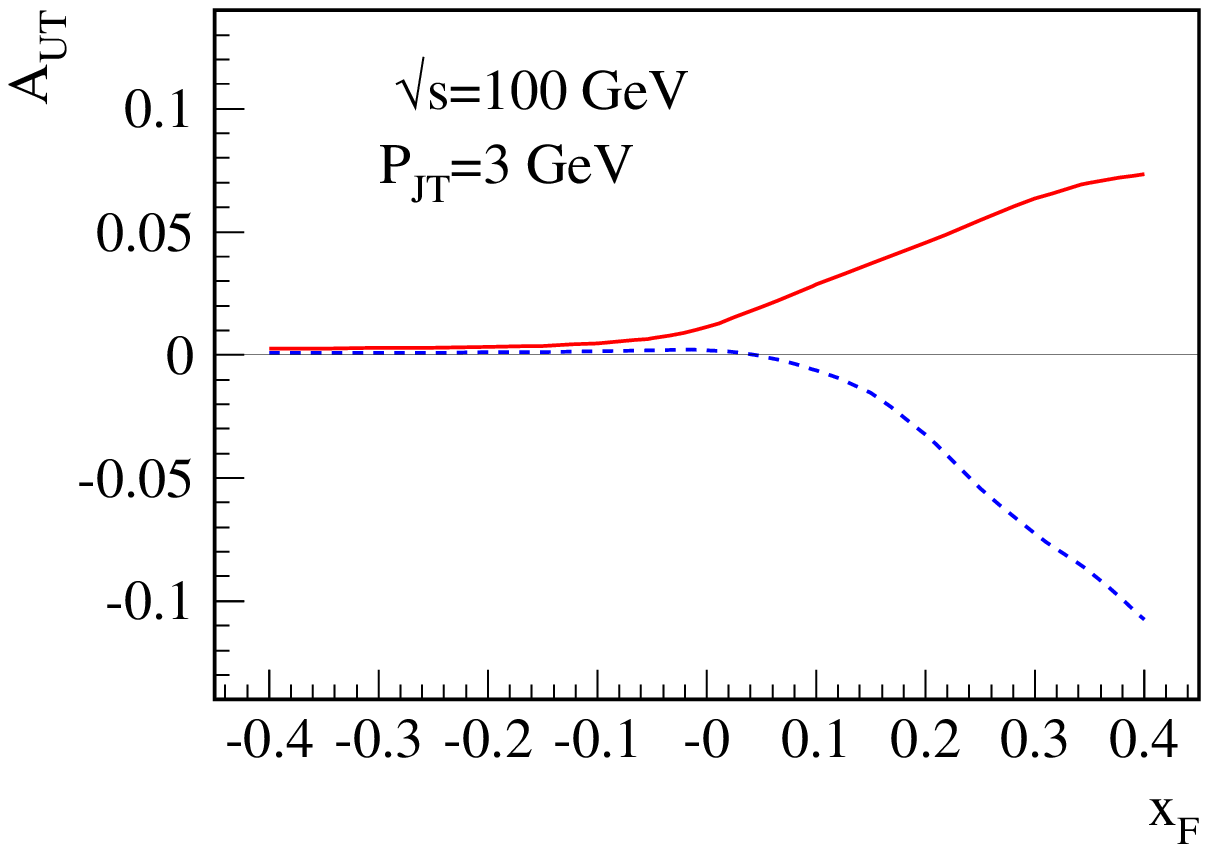}
\caption{$A_{UT}$ as a function of $x_F$ for $\sqrt{s} = 50 \, \textrm{GeV}$ (left)
and $\sqrt{s} = 100 \, \textrm{GeV}$ (right).
The solid line is for $T_F$ taken from the fit of the Sivers function
in~\cite{Anselmino:2008sga} and using the relation in~(\ref{e:qgq_tmd_1}), while the
dashed line is computed with $T_F$ from~\cite{Kouvaris:2006zy,Kang:2011hk}.}
\label{f:aut_1}
\end{center}
\end{figure}

We start by discussing the twist-2 asymmetry $A_{LL}$.
As shown in Fig.~\ref{f:all}, this observable is relatively small (on the percent
level).
It is largest in the backward region (negative $x_F$), and rises with increasing
$P_{JT}$.
Despite the small effect, measuring $A_{LL}$ could provide complementary
information on the quark helicity distributions of the proton.
On the other hand, for the longitudinal double spin asymmetry in
$l \, p \to \textrm{jet} \, X$ one faces the same problems
one has in inclusive DIS: quarks and antiquarks enter with equal weight, and a
flavor separation is hardly possible.
However, if instead one considers $A_{LL}$ for inclusive hadron production
these problems can, in principle, be circumvented like in semi-inclusive DIS.
According to Fig.~\ref{f:all}, $A_{LL}$ clearly increases towards lower
values of $\sqrt{s}$.
Therefore, $A_{LL}$ for $l \, p \to H \, X$ (at $\sqrt{s} < 50 \, \textrm{GeV}$)
should definitely be a very interesting observable for studying the quark helicity
structure of the proton.

\begin{figure}[t]
\begin{center}
\includegraphics[width=7.5cm]{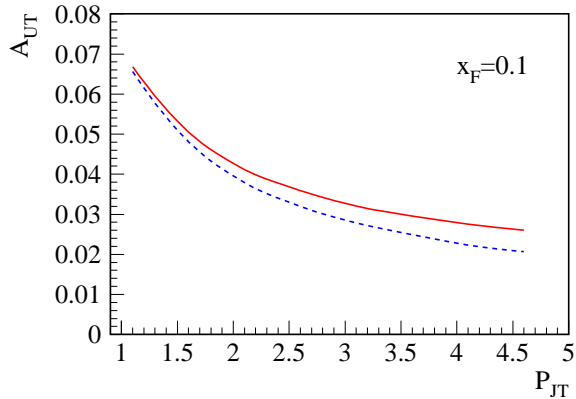}
\caption{$A_{UT}$ as a function of $P_{JT}$ at $x_F=0.1$.
Solid line: $\sqrt{s} = 50 \, \textrm{GeV}$;
dashed line: $\sqrt{s} = 100 \, \textrm{GeV}$.
The matrix element $T_F$ is taken from the fit of the Sivers function
in~\cite{Anselmino:2008sga} and using the relation in~(\ref{e:qgq_tmd_1}).}
\label{f:aut_2}
\end{center}
\end{figure}

Let us now turn to the transverse SSA $A_{UT}$, which is displayed in Fig.~\ref{f:aut_1}
and Fig.~\ref{f:aut_2}.
For both parameterizations we obtain a healthy asymmetry in the forward region,
with effects at the level $5-10 \, \%$.
(Note also that, for the parameterization taken from~\cite{Anselmino:2008sga}, our
numerical results in the collinear approach are similar to those obtained in
Ref.~\cite{Anselmino:2009pn} by using factorization in terms of transverse momentum
dependent parton correlators and the same input for the Sivers function.)
The asymmetry drops with increasing $P_{JT}$ and hardly changes when varying
$\sqrt{s}$ (see Fig.~\ref{f:aut_2}).
The weak energy-dependence of $A_{UT}$ appears because, according to~(\ref{e:master}),
both $\sigma_{UU}$ and $\sigma_{UT}$ have the same hard scattering coefficient.
From Fig.~\ref{f:aut_1} it is obvious that the transverse SSA $A_{UT}$ seems very
promising in order to experimentally constrain the ETQS matrix element $T_F$ and
the Sivers function.
Such a constraint is of utmost importance as the existing parameterizations even
differ in sign~\cite{Kang:2011hk}.
If $P_{JT}$ is sufficiently large, our leading order calculation should give a
reliable estimate of the asymmetry.
One may even be able to check whether $T_F(x,x)$ changes sign as function of $x$,
as has been recently speculated~\cite{Boer:2011fx}.
\begin{figure}[t]
\begin{center}
\includegraphics[width=7.5cm]{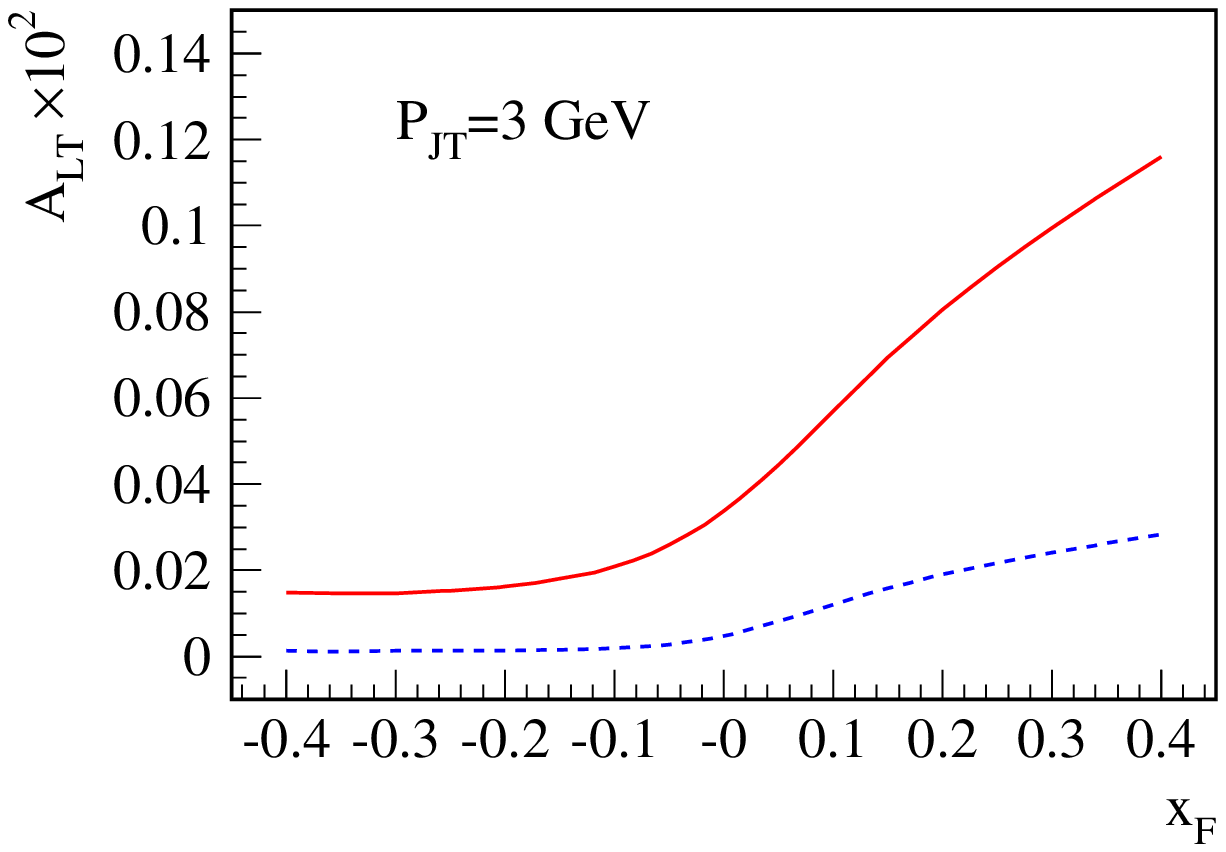}
\hskip 1.0cm
\includegraphics[width=7.5cm]{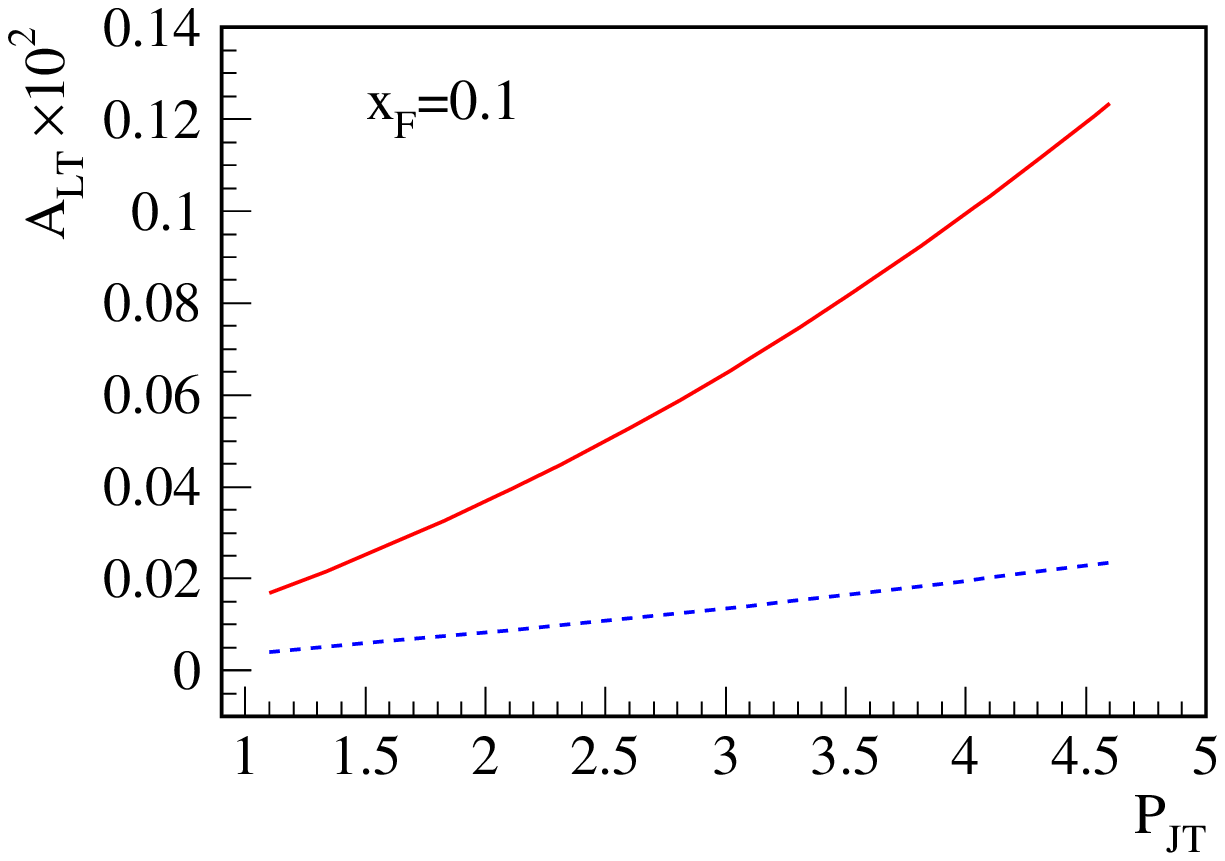}
\caption{$A_{LT}$ as a function of $x_F$ (left) and $P_{JT}$ (right).
Solid line: $\sqrt{s} = 50 \, \textrm{GeV}$;
dashed line: $\sqrt{s} = 100 \, \textrm{GeV}$.
Note that $A_{LT}$ is scaled by a factor of 100.}
\label{f:alt}
\end{center}
\end{figure}

Finally, our numerical estimates for $A_{LT}$ are shown in Fig.~\ref{f:alt}.
This asymmetry is apparently too small to be measured.
The hard scattering coefficient of $\sigma_{LT}$ is the same than the one for
$\sigma_{LL}$.
Therefore, $A_{LT}$ decreases with increasing energy like $A_{LL}$ does.
The main reason for $A_{LT}$ being even much smaller than $A_{LL}$ can be traced
back to the factor $x$ showing up on the RHS of the Wandzura-Wilczek-type
approximation~(\ref{e:appr_2}).
Because of this, an experimental study of $A_{LT}$ can serve as an interesting
check of the relation~(\ref{e:appr_2}), as already pointed out above.
In addition, like in the case of $A_{LL}$, measurable effects for $A_{LT}$ in
hadron production at lower values of $\sqrt{s}$ can be expected.

%
%
\section{Summary and discussion}
\noindent
We have studied a complete set of polarization observables for the process
$l \, p \to \textrm{jet} \, X$.
Neglecting parity violating contributions and transverse polarization of the
lepton one can consider three spin asymmetries: $A_{LL}$, $A_{UT}$, and $A_{LT}$.
We have computed these asymmetries at the level of Born diagrams in collinear
factorization.
Moreover, numerical estimates for typical kinematics of a potential future
Electron Ion Collider have been provided.
(We have explored the {\it cm}-energies $\sqrt{s} = 50 \, \textrm{GeV}$ and
$\sqrt{s} = 100 \, \textrm{GeV}$.)
In the following we summarize our findings and add some discussion:
\begin{itemize}
\item We have discussed in detail how to calculate the process systematically
at higher orders in perturbation theory.
Mainly for two reasons it is important to extend our explicit calculations to
the 1-loop level.
First, one must investigate how stable the various asymmetries are upon inclusion
of NLO-corrections,
in particular, the logarithmically enhanced $\gamma-q$ channel as pointed out in
Sec.~\ref{sec:fac}.
Second, a 1-loop calculation for twist-3 observables, in which derivatives of
quark-gluon-quark correlators (see $d \, T_F / dx$ and $d \, \tilde{g} / dx$
in~(\ref{e:master})) appear, has never been done before.
\item The numerical result for the double spin asymmetry $A_{LL}$ ($A_{LT}$)
is small (tiny) --- on the percent level for $A_{LL}$.
However, in both cases significant effects can be expected for
lower values of $\sqrt{s}$, let's say around $10-20 \, \textrm{GeV}$.
Of course, in this region one cannot perform jet-measurements but has
rather to consider inclusive hadron production.
\item We find that in the forward region the transverse SSA $A_{UT}$ can
become of the order $5-10 \, \%$.
This observable gives a direct handle on the ETQS twist-3 matrix element $T_F$
--- from a theoretical point of view it is one of the simplest observables for
addressing $T_F$ (to leading order neither soft fermion poles, nor tri-gluon
correlations, nor hard gluon poles contribute) ---  and also, by means
of~(\ref{e:qgq_tmd_1}), to the transverse momentum dependent Sivers function
$f_{1T}^\perp$.
Given the fact that at present even the sign of $T_F$ is unclear~\cite{Kang:2011hk},
experimental information on this observable would be very valuable.
\end{itemize}
In general, we believe that there is sufficient justification for further
exploring the potential of high-energy lepton nucleon scattering with an
unidentified final state lepton.
Such type of reaction may also constitute an interesting part of the
physics program at a future Electron Ion Collider.
\\[0.5cm]
%
%
\noindent
{\bf Acknowledgments:}
We thank Naomi Makins, Werner Vogelsang, and Feng Yuan for helpful discussions.
We are grateful to RIKEN, Brookhaven National Laboratory, and the U.S.~Department
of Energy (Contract No.~DE-AC02-98CH10886) for providing the facilities essential
for the completion of this work.
A.M. acknowledges the support of the NSF under Grant No.~PHY-0855501.

%
%

\end{document}